\documentclass[%
 reprint,
superscriptaddress,
 amsmath,amssymb,
 aps,
pra,
]{revtex4-2}
\usepackage{graphicx} 
\usepackage{dcolumn}
\usepackage{bm}

\usepackage{braket}
\usepackage{color}

\begin{document}
\title{Improving the efficiency of quantum annealing with controlled diagonal catalysts}
\author{Tomohiro Hattori$^1$}
\affiliation{$^1$ Graduate School of Science and Technology, Keio University, 3-14-1 Hiyoshi, Kohoku-ku, Yokohama-shi, Kanagawa 223-8522, Japan}
\author{Shu Tanaka$^{1,2,3,4}$\thanks{shu.tanaka@appi.keio.ac.jp}}
\affiliation{$^2$ Department of Applied Physics and Physico-Informatics, Keio University, 3-14-1 Hiyoshi, Kohoku-ku, Yokohama-shi, Kanagawa 223-8522, Japan \\
$^3$ Keio University Sustainable Quantum Artificial Intelligence Center (KSQAIC), Keio University, Tokyo 108-8345, Japan \\
$^4$ Human Biology-Microbiome-Quantum Research Center (WPI-Bio2Q), Keio University, Tokyo 108-8345, Japan
}

\date{\today}
\begin{abstract}
Quantum annealing is a promising algorithm for solving combinatorial optimization problems.
It searches for the ground state of the Ising model, which corresponds to the optimal solution of a given combinatorial optimization problem.
The guiding principle of quantum annealing is the adiabatic theorem in quantum mechanics, which guarantees that a system remains in the ground state of its Hamiltonian if the time evolution is sufficiently slow.
According to the adiabatic theorem, the annealing time required for quantum annealing to satisfy the adiabaticity scales inversely proportional to the square of the minimum energy gap between the ground state and the first excited state during time evolution.
As a result, finding the ground state becomes significantly more difficult when the energy gap is small, creating a major bottleneck in quantum annealing.
Expanding the energy gap is one strategy for improving the performance of quantum annealing; however, its implementation in actual hardware remains difficult.
This study proposes a method for efficiently solving instances with small energy gaps by introducing additional local terms to the Hamiltonian and exploiting the diabatic transition remaining in the small energy gap.
The proposed method achieves an approximate quadratic speedup of the exponential scaling exponent in time to solution compared to the conventional quantum annealing.
In addition, we investigate the transferability of the parameters obtained with the proposed method.
\end{abstract}
\maketitle

\section{\label{sec: introduction}Introduction}
Quantum annealing (QA)~\cite{kadowaki1998quantum,farhi2000quantum,mcgeoch2014adiabatic,tanaka2017quantum,albash2018adiabatic,chakrabarti2023quantum} is a quantum algorithm that has the potential to solve combinatorial optimization problems more efficiently than existing classical methods.
In QA, the solution of a combinatorial optimization problem corresponds to the ground state of the Ising model, which QA is designed to obtain~\cite{lucas2014ising, tanahashi2019application}.

The first commercial quantum annealer~\cite{johnson2011quantum} significantly expanded the range of applications for QA.
Recently, QA has been applied to various complex problems, including 
combinatorial optimizations with integer variables~\cite{chancellor2019domain, okada2019efficient, chen2021performance, tamura2021performance, seki2022black, kikuchi2024performance}, large-scale combinatorial optimizations~\cite{karmi2017boosting, karimi2017effective, okada2019improving, irie2021hybrid, Atobe_2022, kikuchi2023hybrid, hattori2025advantages}, and black-box optimizations~\cite{FMA2020, photonic_laser2022, matsumori2022application, FMA2022, FMA2023, kim2024quantum, xiao2024application, endo2025function, tamura2025black}.
In addition, QA has been applied to the fields of physical simulation and computer-aided engineering~\cite{harris2018phase, king2018observation, honda2024development, takagi2024implementation, xu2025quantum}.

The guiding principle of QA is the adiabatic theorem in quantum mechanics~\cite{kato1950adiabatic}, which states that a system remains in its ground state during the evolution of a time-dependent Hamiltonian, provided that the evolution is sufficiently slow.
The annealing time required to perform QA while maintaining adiabaticity is inversely proportional to the square of the minimum energy gap between the ground state and the first excited state during the annealing process.

However, the hardware limitations of quantum annealers, such as decoherence, size restrictions, and energy range restrictions, limit their practical applicability.
The parameter tuning proposed in recent studies is crucial in overcoming these hardware limitations, as they affect QA performance significantly~\cite{raymond2025quantum, Braida2024rescalingparameter, kikuchi2023hybrid, hattori2025advantages, hattori2025impact}.
The annealing time is restricted due to the decoherence in the hardware.
As a result, combinatorial optimization problems that require long annealing times, particularly when the QA Hamiltonian exhibits a small energy gap, are especially difficult to solve on current hardware.

A well-known bottleneck of QA is the occurrence of perturbative crossings and quantum first-order phase transitions during the annealing process.
When perturbative crossings or quantum first-order phase transitions occur, the energy gap decreases exponentially with system size, as has been analytically proven~\cite{altshuler2010anderson}.
Numerous studies have been conducted to mitigate these bottlenecks~\cite{somma2012quantum, seki2012quantum, susa2018exponential, susa2018quantum, adame2020inhomogeneous, albash2021diagonal, feinstein2024effects, ghosh2024exponential}.

One promising approach to overcoming these bottlenecks is to expand the energy gap.
Previous studies have shown that adding an extra term, a catalyst, is crucial in achieving this expansion.
A well-known example is the introduction of antiferromagnetic quantum fluctuations in the fully connected $p$-spin model~\cite{seki2012quantum}, leading to an exponential increase in the energy gap.
In addition, $XX$ catalysts have been shown to improve the scaling of the energy gap in the maximum weighted independent set problem, a well-known toy model that exhibits perturbative crossings~\cite{feinstein2024effects, ghosh2024exponential}.
Moreover, under certain conditions, diagonal catalyst terms in the computational basis have been found to exponentially improve the energy gap scaling in the fully connected $p$-spin model~\cite{albash2021diagonal}.

Implementing these approaches in existing QA hardware remains difficult due to the quadratic terms of these catalysts.
Quadratic terms, such as $ZZ$-interactions and $XX$-interactions used as catalyst terms, are not yet realizable in current hardware.
Since the diagonal catalysts consist only of linear terms, it is more amenable to implementation on hardware.
However, this study points out that optimization becomes difficult when the Hamming distance between the global optimum and local optima is large~\cite{albash2021diagonal}.
Therefore, exploring the potential of linear-term catalysts is essential for practical implementations.

Another practical approach in a quantum annealer is optimizing the annealing schedule.
Optimization of the annealing schedule can reduce the time required to obtain optimal solutions~\cite{cote2023diabatic, finvzgar2024designing}.
In a previous study, annealing-schedule optimization improved the QA performance for the frustrated ring model, which exhibits perturbative crossing~\cite{cote2023diabatic}.
Also, the previous study proposes the optimization method using Bayesian optimization on a neutral-atom quantum processor for both QA and reverse annealing~\cite{finvzgar2024designing}.
An optimized schedule significantly accelerates the QA process compared to a linear schedule.

There also exist parameter-optimization approaches based on quantum computing by coherent cooling~\cite{arisoy2021few, feng2022quantum} and quantum walks~\cite{imparato2024thermodynamic}. 
Since the system dynamics in these methods are governed by operating principles different from the adiabatic theorem, they are expected to be effective for problem instances that are challenging for QA.
However, practical implementation on current hardware remains challenging due to the effects of intermediate-state quenches and decoherence.

When performing QA with an optimized annealing schedule, the evolution of the quantum system does not strictly adhere to the adiabatic theorem in quantum mechanics.
Instead, the system leverages diabatic transitions and temporarily occupies excited states during annealing, driven by the optimized annealing schedule.
Consequently, the proposed method leverages diabatic dynamics to enhance performance.
However, a detailed theoretical understanding of the underlying mechanism remains incomplete.

The annealing-schedule optimization method also faces practical restrictions arising from hardware limitations.
Implementing an optimized annealing schedule that involves quadratic terms in the Ising Hamiltonian is particularly difficult since errors arise from the coupled spins, and the quadratic term makes a significant contribution to error accumulation~\cite{D-Wave_error}.
Additional physical qubits are necessary when embedding a densely connected Ising model onto hardware with restricted connectivity.
For instance, the minor embedding~\cite{choi2008minor, choi2011minor} technique introduces redundant qubits to express the original dense problem graph onto the hardware topology.
These interactions between the redundant qubits are strong enough to align these qubits in the same direction.

When optimizing the annealing schedule, the effects of couplers between redundant qubits must be considered.
Nevertheless, the number of couplers and qubits grows quadratically with the size of the Ising model, and precise control of the annealing schedule is required.
However, the precise control of the complex terms is difficult in current hardware.
From this perspective, the scalability of the algorithm is inherently limited.

This study proposes optimizing the annealing schedule using only linear terms of $z$ fields.
Specifically, we introduce additional longitudinal magnetic fields into the QA Hamiltonian.
The schedule of these additional longitudinal magnetic fields is optimized variationally, while the annealing schedules of other terms are fixed.
The proposed approach is highly practical for hardware implementation because the additional local $z$-fields can be realized through scheduling time-dependent linear coefficients with \verb|h_gain_schedule|~\cite{D-Wave_solver_para}.
Implementing linear $z$-field terms is significantly easier than incorporating quadratic $z$-terms, further improving the feasibility.
This makes it possible to harness the benefits of annealing schedule optimization solely with linear terms, without relying on the complex control of quadratic terms, whose implementation remains limited.
Numerical experiments demonstrate that our proposed method solves Maximum Weighted Independent Set (MWIS) problems faster than conventional QA with a linear annealing schedule.

Our results show that optimizing the local $z$-field schedule can improve optimization performance by enabling a hybrid dynamical behavior: the system evolves adiabatically in regions where the energy gap is sufficiently large, while selectively exploiting nonadiabatic transitions only in regions where strictly adiabatic evolution along a single path becomes difficult.

The remainder of this paper is organized as follows.
Section~\ref{sec: settings} describes the QA and its evaluation metric, MWIS settings, and the proposed method.
Section~\ref{sec: result} analyzes the performance of the proposed method via numerical simulations, focusing on time to solution (TTS), population dynamics of the proposed method, the properties of the optimized annealing schedules, and their transferability.
Finally, Section~\ref{sec: discussion} presents our conclusions and outlines future directions. 

\section{\label{sec: settings}Settings}
We review QA and MWIS, which we use as a benchmark problem with perturbative crossings.
We then introduce the proposed method, which utilizes local magnetic fields as a catalyst.

\subsection{\label{subsec: QA}Quantum annealing}
The Hamiltonian of QA is expressed as follows:  
\begin{align}
    \label{eq:H}
    \mathcal{H}(t) = A(t)\mathcal{H}_\mathrm{q} + B(t)\mathcal{H}_\mathrm{p},\quad 0\leq t\leq \tau,
\end{align}
where $A(t)$ and $B(t)$ are the annealing schedules which satisfy $A(0)\gg B(0)$ and $A(\tau) \ll B(\tau)$.
$\mathcal{H}_\mathrm{q}$ is the transverse magnetic field.
$\mathcal{H}_\mathrm{p}$ is the Ising model whose ground state corresponds to the solution of the combinatorial optimization problem.
Thus, finding the ground state of $\mathcal{H}_\mathrm{p}$ is equivalent to solving the combinatorial optimization problem.
The quantum system evolves following the Schr\"{o}dinger equation, which is denoted as follows:
\begin{align}
    \label{eq: schrodinger_equation}
    \mathrm{i}\frac{\partial }{\partial t}\ket{\psi(t)} = \mathcal{H}
    (t)\ket{\psi(t)}.
\end{align}
Here, Planck units are used.

According to the adiabatic condition~\cite{tanaka2017quantum}, the annealing time required to remain in the instantaneous ground state scales as the inverse square of the minimum energy gap during QA.
Consequently, the energy gap is widely used as a key metric of QA performance.
The energy gap $\Delta(t)$ between the ground state $\ket{\phi_0(t)}$ and the first excited state $\ket{\phi_1(t)}$ is defined as
\begin{align}
    \label{eq: energy_gap}
    \Delta(t)=E_1(t)-E_0(t),
\end{align}
where $E_1(t)$ and $E_0(t)$ denote the eigenenergies of $\ket{\phi_1(t)}$ and $\ket{\phi_0(t)}$, respectively.

\subsection{\label{subsec: MWIS}Perturbative crossing in Maximum Weighted Independent Set}
\begin{figure}[t]
    \centering
    \includegraphics[scale=0.4]{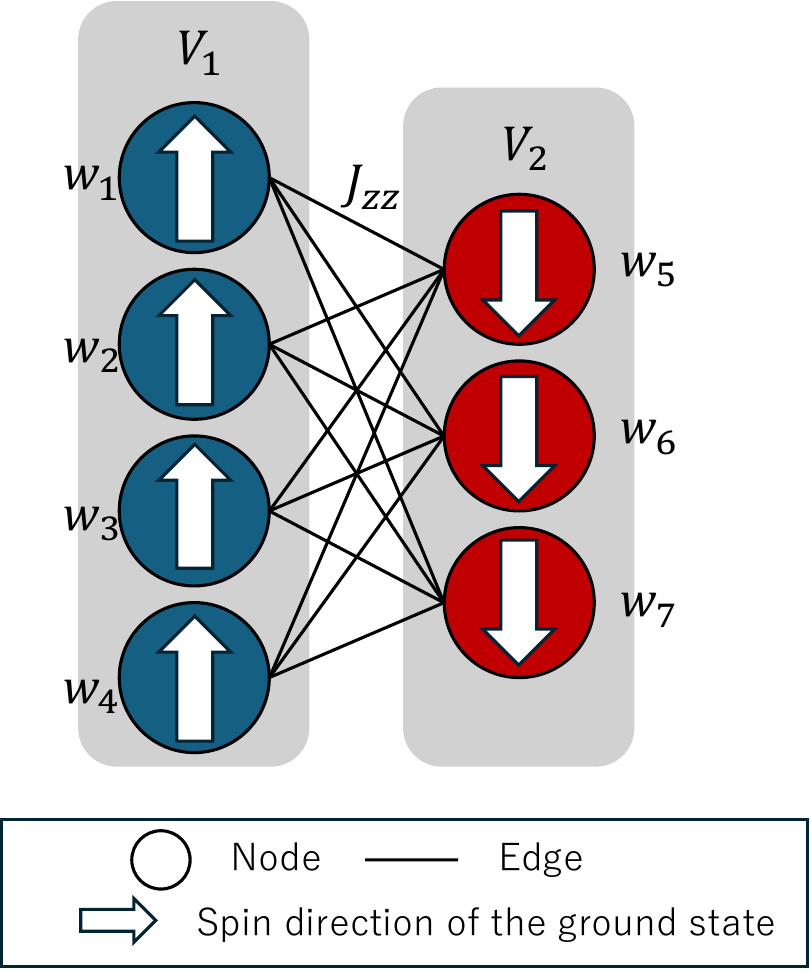}
    \caption{Conceptual diagram of the Ising model for MWIS on $K_{4,3}$.
    The circles represent the vertices of $K_{4,3}$, and the black lines indicate the edges.
    The graph consists of two disjoint subsets, shown as dark blue and red circles.
    Each spin on the nodes of the graph has an associated weight denoted by $w_i$.
    The white arrows represent the spin directions in the ground state of the Ising model.
    All edge interactions in $K_{4,3}$ have the same value $J_{zz}$.
    }
    \label{fig: MWIS}
\end{figure}
\begin{figure*}[t]
    \centering
    \includegraphics[scale=1]{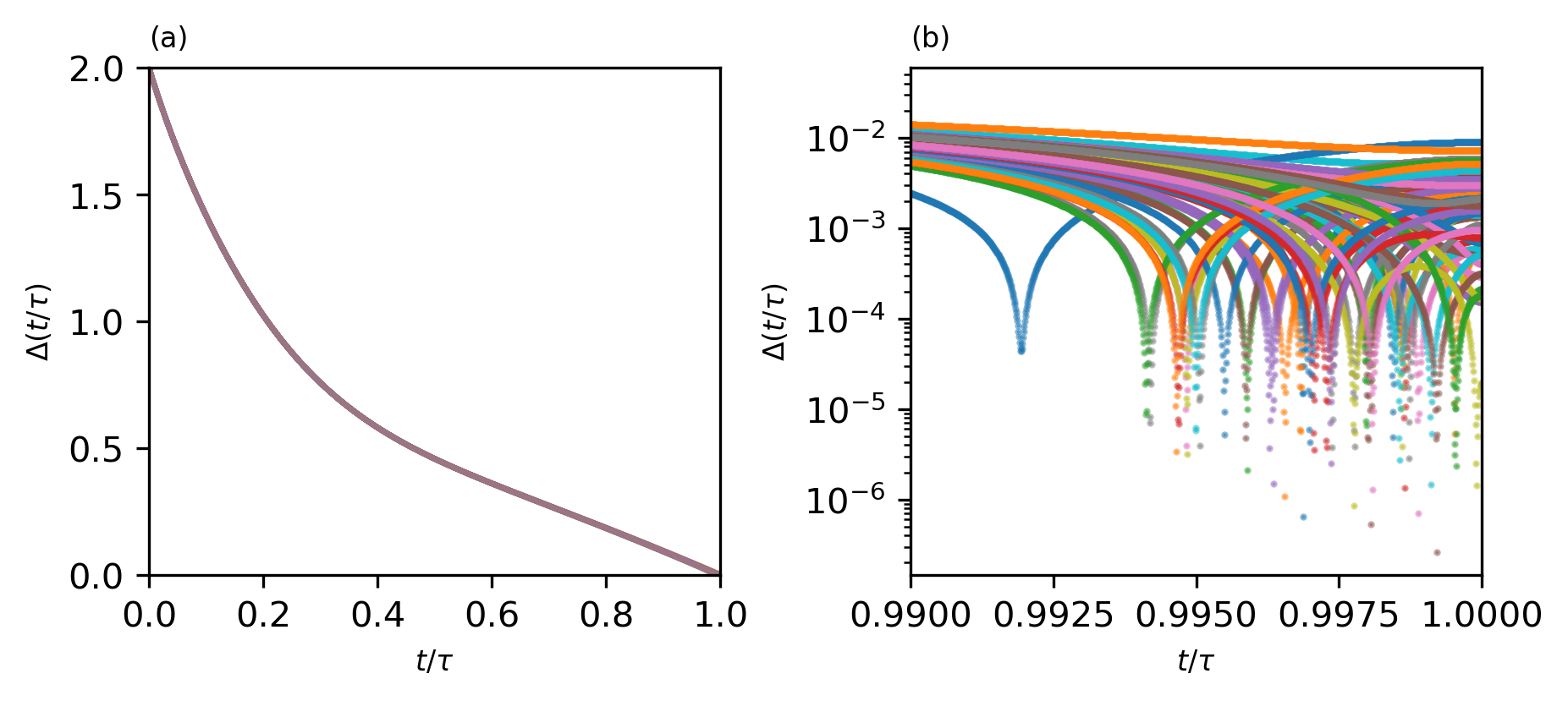}
    \caption{(a) Energy gap $\Delta(t)$ between the ground state and the first excited state as a function of $t$ for MWIS instances with an $N$-Hamming-distance configuration.
    Since the energy difference between the ground state and the first excited state in the problem Hamiltonian is extremely small, the energy gap $\Delta(t)$ closes near the end of the annealing process.
    (b) Enlarged view of $\Delta(t)$ in the range $0.99 \leq t/\tau \leq 1$ on a logarithmic scale, showing the rapid closure of the gap near the final annealing time.
    }
    \label{fig: eg_MWIS}
\end{figure*}

Under certain conditions, the energy gap closes near the end of the QA, known as the perturbative crossing.
The perturbative crossing occurs when the problem Hamiltonian $\mathcal{H}_\mathrm{p}$ has a very small energy gap between the ground and first excited states, and when these states differ by a large Hamming distance~\cite{altshuler2010anderson}.

MWIS on a complete bipartite graph $G(V_1 \cup V_2, E)$ is one of the combinatorial optimization problems, where $V_1 \cup V_2$ and $E$ are the sets of vertices and edges, respectively. 
The vertex set is partitioned into two disjoint subsets $V_1$ and $V_2$ (i.e., $V_1 \cap V_2 = \emptyset$), with no edges within each subset and all-to-all edges between subsets.
If $|V_1|=m$ and $|V_2|=n$, the complete bipartite graph is denoted by $K_{m,n}$, where $|\cdot|$ denotes the number of elements in the set.
We set $m=n+1$ to set different $c_i$ for vertices in different subsets following the previous studies~\cite{feinstein2024effects, ghosh2024exponential}.
The number of nodes is $N$, which satisfies $N=m+n$.

MWIS on the complete bipartite graph is formulated as the following Ising Hamiltonian:
\begin{align}
    \label{eq: MWIS}
    \mathcal{H}_\mathrm{p} =\sum_{i\in V_1 \cup V_2}\left(c_iJ_{zz}-w_i \right) \sigma_i^z +\sum_{(i,j)\in E}J_{zz}\sigma_i^z\sigma_j^z,
\end{align}
where $J_{zz}$, $c_i$, and $w_i$ are the weight of edges, the number of edges connected to vertex $i$, and the weight on vertex $i$, respectively, and $\sigma_i^z$ denotes the $z$-component of the Pauli matrix on vertex $i$.

We generated $500$ MWIS instances under conditions leading to perturbative crossings.
All interactions between the subsets $V_1$ and $V_2$ are set to be antiferromagnetic with $J_{zz}=1$, thereby taking $J_{zz}$ as the unit of energy.
Specifically, each instance was constructed with very small weights $w_i$, randomly assigned from a uniform distribution in the $[-0.005/N, 0.005/N]$ interval. 
Due to these small values of $w_i$, the energy difference between the ground state and the first excited state becomes extremely small, leading to perturbative crossings. 
Dividing by $N$ preserves the energy scale of the magnetic field. This prevents the total sum of $w_i$ from becoming excessively large and helps maintain the conditions under which the energy gap remains small, even for random choices of $w_i$.
Under these conditions, the ground state and the first excited state are predominantly determined by subtle differences in the summation term $\sum_{i\in V_1 \cup V_2}w_is_i$, where $s_i$ is the Ising variable on vertex $i$.

Figure~\ref{fig: MWIS} shows an example of the ground state configuration of the Ising formulation of MWIS on $K_{4,3}$.
Since $w_i$ is negligible compared to $J_{zz}$, the antiferromagnetic interactions dominate the spin configuration, outweighing the effect of the local magnetic fields.
Therefore, the spins in $V_1$ and $V_2$ align in opposite directions in the ground state, as shown in Fig.~\ref{fig: MWIS}.

Among $500$ MWIS instances, we observed two types: those in which the ground and first excited states differ by a Hamming distance of $1$, and those in which they differ by a Hamming distance of $N$.
A useful metric for assessing the difficulty of classical optimization is the ratio between the total accumulated energy differences across all two-state combinations and those involving states at Hamming distance 1~\cite{weinberger1990correlated, verma2022penalty}.
Instances with only small energy differences at Hamming distance 1 are relatively easy to solve.
Because classical algorithms with $O(N)$ complexity, such as single-spin-flip methods, can efficiently handle these cases, we focus instead on instances with Hamming distance $N$.
We further selected instances in which the energy gap between the ground and first excited states exhibits a pronounced valley around $t/\tau\simeq 1$. 
Therefore, MWIS on a complete bipartite graph with the Hamming distance $N$ between the ground state and the first excited state is a crucial benchmark problem.

Figure~\ref{fig: eg_MWIS} shows the energy gap $\Delta(t)$ of $\mathcal{H}(t)$ for the selected MWIS instances.
$\Delta(t)$ closes near the end of the annealing time in all instances, as shown in Fig.~\ref{fig: eg_MWIS}~(a).
Figure~\ref{fig: eg_MWIS}~(b) magnifies the region around the final annealing time, where $\Delta(t)$ closes sharply in most instances.
For these cases, conventional QA has difficulty reaching the ground state at the final time, since the small gap makes the system susceptible to excitations into higher energy levels.
We therefore examined the performance of the proposed method for such hard MWIS instances.

We also examined the effects of the proposed method in the Sherrington--Kirkpatrick (SK) spin-glass problem in Appendix~\ref{sec: SK}.
\subsection{Proposed method}
Previous studies have explored various types of catalysts~\cite{seki2012quantum, feinstein2024effects, ghosh2024exponential, albash2021diagonal}.
Such catalysts can yield exponential speedups of QA for specific problems.
However, most proposed catalysts are nonlocal and therefore difficult to implement on existing QA hardware.
In this study, we introduce a linear-term catalyst to investigate the potential of the local catalysts.
The Hamiltonian of the QA with the catalyst is given by 
\begin{align}
    \label{eq: H_catalyst}
    \mathcal{H}(t) = A(t)\mathcal{H}_\mathrm{q} +B(t)\mathcal{H}_\mathrm{p} +C(t)\mathcal{H}_\mathrm{catalyst},
\end{align}
where $C(t)$ is the annealing schedule of the catalyst restricted with $C(0)=C(\tau)=0$, the $\mathcal{H}_\mathrm{catalyst}$ is the diagonal catalyst, which is expressed as follows:
\begin{align}
    \label{eq: zcatalyst}
    \mathcal{H}_\mathrm{catalyst} = -\sum_{i=1}^N\sigma_i^z.
\end{align}
Since $\mathcal{H}_\mathrm{catalyst}$ contains only the linear term, the optimized schedule is easier to implement in hardware.
Indeed, such a restricted diagonal catalyst in hardware can be realized using individual scheduling of local magnetic fields~\cite{D-Wave_solver_para}.

$C(t)$ is variationally optimized in the proposed method.
To implement a gradient-based optimization, we employed optimal control theory~\cite{todorov2006optimal}, which has also been applied in previous studies~\cite{brady2021optimal, shirai2024post}.
The framework enables us to compute the gradient of the final-time expected energy with respect to $C(t)$ by making use of the quantum state at each time $t$.

First, we define the control problem as the expected energy at the final time, which is denoted as follows:
\begin{align}
    \label{eq: control_problem}
    \mathcal{J}=\bra{\psi(\tau)}\mathcal{H}_\mathrm{p}\ket{\psi(\tau)},
\end{align}
where $\ket{\psi(t)}$ is the quantum state at the time $t$.
To obtain the optimal $C(t)$ that minimizes the final-time expected energy, we compute the functional derivative of $\mathcal{J}$ with respect to $C(t)$:
\begin{align}
    \label{eq: gradient_of_control_problem}
    \frac{\partial\mathcal{J}}{\partial C(t)} = 2\mathrm{Im}\left(\bra{k(t)}\mathcal{H}_\mathrm{catalyst}\ket{\psi(t)}\right),
\end{align}
where $\mathrm{Im}(\cdot)$ denotes the imaginary part, $\ket{k(\tau)}$ satisfies $\ket{k(\tau)}=\mathcal{H}_\mathrm{p}\ket{\psi(\tau)}$, and $\ket{k(t)}$ is obtained by numerically solving the Schr\"{o}dinger equation by using time-reversal properties of the Schr\"{o}dinger equation.
The derivation of the Eq.~\eqref{eq: gradient_of_control_problem} is shown in the appendix~\ref{sec: Derivation_optsche}.
In practice, a more realistic strategy is to employ variational optimization methods as demonstrated in a previous study~\cite{cote2023diabatic}.

Using the gradient derived in Eq.~\eqref{eq: gradient_of_control_problem}, we iteratively update annealing schedule $C(t)$ as follows:
\begin{align}
    \label{eq: renew_annealing_schedule}
    C^{(p+1)}(t)\leftarrow C^{(p)}(t) -\eta \frac{\partial\mathcal{J}}{\partial C^{(p)}(t)},
\end{align}
where $C^{(p)}(t)$ denotes the annealing schedule after $p$ optimization steps and $\eta$ is the learning rate.
Since previous studies suggest that the schedule optimization method exhibits weak sensitivity to initial parameters and typically converges to optimal or near-optimal solutions~\cite{brady2021optimal}, we fixed the initial parameters $C(t)=0~\forall t$.
The initial schedule design helps avoid performance degradation compared to conventional QA.
In this study, we used $1000$ iterations and $\eta=0.01$.

In this study, the annealing schedules $A(t)$ and $B(t)$ are set as $A(t)=1-t/\tau$ and $B(t)=t/\tau$, respectively, while the optimized $C(t)$ is used.
The time evolution following Eq.~\eqref{eq: schrodinger_equation} with $\mathcal{H}(t)$ written in Eq.~\eqref{eq: H_catalyst} is performed in our proposed method.

\section{\label{sec: result}Results}
We numerically examined both the performance of the proposed method and the transferability of its optimized schedule.
The QA simulation is carried out by solving the Schr\"{o}dinger equation with QuTiP~\cite{QuTiP1, QuTiP2}.
In this study, we set $\tau=512$, which is the long annealing time compared to $J_{zz}=1$.
However, due to a small energy gap, the QA with a linear schedule fails to reach the ground state at the final annealing time with $\tau=512$.

\subsection{\label{subsec: TTS}Performance of the proposed method}
This study aims to obtain the optimal solution faster than the conventional method.
There is, however, a trade-off between achieving a high ground-state probability in a single long run and performing multiple shorter runs with a lower single-run ground-state probability.
time to solution (TTS) quantifies this trade-off by measuring the time required to achieve the desired ground-state probability $p_{\rm d}$ (typically $p_{\rm d}=0.99$), which can reflect the trade-off~\cite{boixo2014evidence,ronnow2014defining}.
Since TTS depends on the annealing time, we denote it as TTS as $T(\tau)$, which is given by
\begin{align}
    \label{eq: TTS}
    T(\tau) = \tau\frac{\ln(1-p_{\rm d})}{\ln(1-p(\tau))},
\end{align}
where $p(\tau)$ is the ground-state probability obtained by QA.
In this study, we set $p_{\rm d}=0.99$.
The dependence of the annealing time is shown in Appendix~\ref{sec: annealing_time_dependence}.

Figure~\ref{fig: TTS_sizescaling} shows the size scaling of TTS.
TTSs of the QA with a linear schedule and the proposed method scale exponentially with $N$. 
However, the proposed method yields consistently shorter TTS than the QA with a linear schedule. 
Table~\ref{tab: TTS_sizescaling} summarizes the fitting functions obtained using the least-squares method.
As shown in the table, the scaling coefficient for the proposed method ($0.46$) is roughly half that for linear-schedule QA ($0.85$), corresponding to an approximate quadratic improvement in the exponential scaling exponent.
These results demonstrate that even a local catalyst has the potential to enhance QA performance.

\begin{figure}[t]
    \centering
    \includegraphics[scale=1,clip]{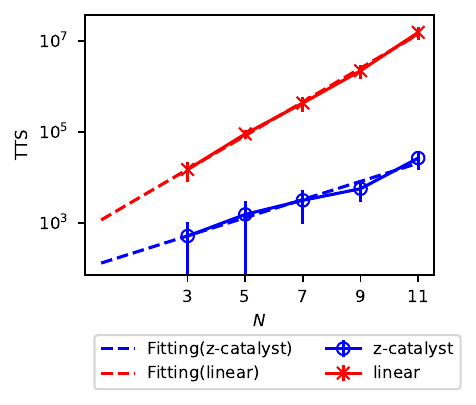}
    \caption{Size scaling of TTS for hard MWIS instances with $\tau=512$.
    The vertical axis is logarithmic.
    Points indicate mean TTS values, error bars denote standard deviations, and dashed lines show least-squares fits (see Table~\ref{tab: TTS_sizescaling}).}
    \label{fig: TTS_sizescaling}
\end{figure}
\begin{table}[t]
    \caption{Fitting functions for TTS obtained using the least-squares method. 
    $C_1$ and $C_2$ represent constant parameters.}
    \centering
    \begin{tabular*}{\columnwidth}{@{\extracolsep{\fill}}lc}
        \hline
        \hline
        Method & Fitting function\\ \hline
        Linear schedule:&
        $T = C_1\exp(0.85N\pm0.0195)$ \\ \hline 
        Proposed method:&
        $T = C_2\exp(0.46N\pm0.0440)$\\ \hline\hline
    \end{tabular*}
    \label{tab: TTS_sizescaling}
\end{table}

\subsection{\label{subsec: GSP}Energy level population}
The guiding principle of QA is the adiabatic theorem.
Therefore, many previous studies aim to follow the adiabatic path~\cite{albash2018adiabatic, campbell2015shortcut, chen2010fast,hatomura2017shortcuts, hatomura2024shortcuts}.
However, the proposed method aims to minimize the expected energy at the final time, as given by Eq.~\eqref{eq: control_problem}.
Hence, the path to the ground state in the proposed method is not necessarily adiabatic.
The diabatic path has the potential to surpass the adiabatic QA.

To investigate the dynamics of the proposed method, we numerically simulated the 
energy-level population during QA.
The population at the $l$th eigenstate is denoted as follows:
\begin{align}
    \label{eq: energy_population}
    P_l(t) = |\braket{\phi_l(t)|\psi(t)}|^2,
\end{align}
where $\ket{\phi_l(t)}$ is the $l$th eigenstate of $\mathcal{H}(t)$.

\begin{figure}
    \centering
    \includegraphics[scale=1.0,clip]{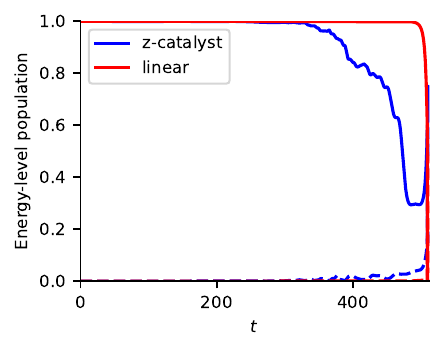}
    \caption{Populations of the ground and first excited states during the time evolution for a hard MWIS instance with $N=11$.
    The annealing time is set to $\tau=512$.
    The blue and red solid lines show the ground-state populations for the proposed method and the linear-schedule QA, respectively, while the dashed lines indicate the corresponding first-excited-state populations.
    The color distinctions are the same as before.
}
    \label{fig: energy_population}
\end{figure}

Figure~\ref{fig: energy_population} shows the result of the ground state population and the first excited state population at time $t$ for the specific instance with $N=11$.
For QA with a linear schedule, the dynamics initially follow the ground state but deviate near the end of the annealing process as the energy gap narrows.
On the other hand, the dynamics of the proposed method deviate from the ground state earlier than those of QA with a linear schedule, and the ground state population decreases as time evolves.
Finally, the population of the ground state increases when the energy gap becomes minimal.
Other hard MWIS instances exhibit similar energy-level population behavior.
These dynamics differ from the adiabatic path and correspond to diabatic QA, as previously discussed in Ref.~\cite{cote2023diabatic}.
The proposed method realizes the diabatic QA by optimizing the schedule of only the linear terms.
This indicates that individual manipulation of linear terms is sufficient to accelerate the search for hard instances, while fine-tuning quadratic terms is unnecessary.
The proposed method can obtain the ground state, which the QA with a linear schedule cannot do, even if the annealing time increases.

\subsection{\label{subsec: transferability}Optimized schedule transferability}
\begin{figure}
    \centering
    \includegraphics[scale=1,clip]{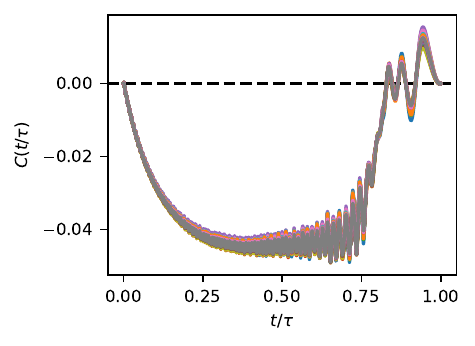}
    \caption{Optimized schedules for hard MWIS instances with $N=11$ at annealing time $\tau=512$.
    The black dashed line represents the zero reference, and different colors correspond to different MWIS instances.
    }
    \label{fig: transferability_bet_MWISs}
\end{figure}
\begin{figure}
    \centering
    \includegraphics[scale=1,clip]{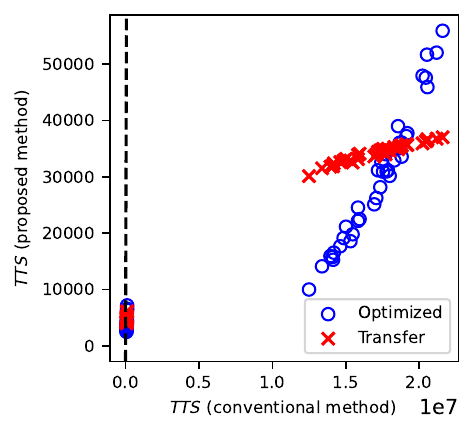}
    \caption{Comparison of TTS between QA with a linear schedule and the proposed method for hard MWIS instances with $N=11$ at $\tau=512$.
    The blue circles represent TTS of the proposed method with an individually optimized additional schedule, while the red crosses represent TTS of the proposed method with an additional schedule transferred from an arbitrarily selected hard MWIS instance.
    The dashed black line represents the equal TTS between the proposed method and the QA with a linear schedule.
    }
    \label{fig: transferability_bet_MWISs_TTS}
\end{figure}
\begin{figure}
    \centering
    \includegraphics[scale=0.7,clip]{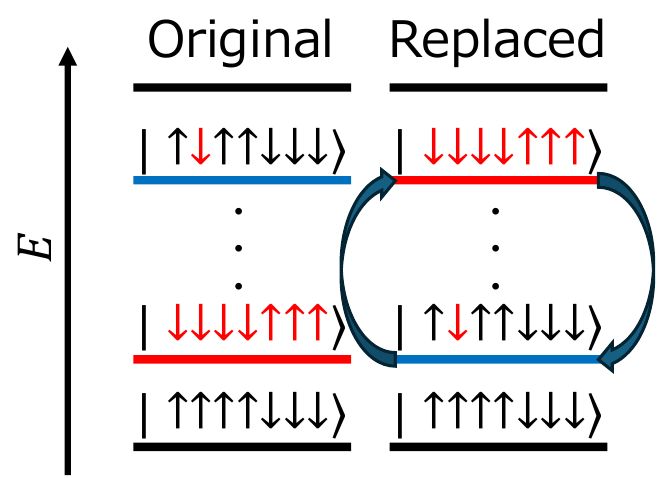}
    \caption{Schematic diagrams of the energy spectrum of the original Hamiltonian and the replaced energy-level Hamiltonian.
    The Hamming distance between the original Hamiltonian's ground state and the first excited state is $N$.
    The Hamming distance between the ground state and the first excited state of the replaced energy-level Hamiltonian is $1$, while preserving the same energy levels, since only the spin configuration is replaced.
    }
    \label{fig: replaced_Hamiltonian}
\end{figure}
\begin{figure}
    \centering
    \includegraphics[scale=1,clip]{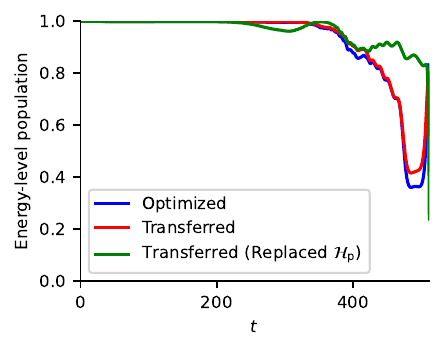}
    \caption{Ground-state population during the time evolution for a hard MWIS instance with $N=11$.
    Similar behavior is observed in other hard MWIS instances.
    The blue solid line denotes the ground-state population of the proposed method, the red solid line that of the proposed method with a transferred schedule, and the green solid line that of the replaced energy-level Hamiltonian, where the Hamming distance between the ground and first excited states is set to $1$.
    }
    \label{fig: transferability_replaced_Hamiltonian}
\end{figure}
The proposed method in Section~\ref{subsec: TTS} achieves a significantly shorter TTS than QA with a linear schedule.
However, it incurs additional optimization costs to obtain this improvement.
Therefore, considering the time required to obtain an optimized schedule, the proposed method is not necessarily superior to the conventional approach for overall computational efficiency.
Thus, eliminating the schedule optimization cost is crucial for the practical applicability of the proposed method.

In previous studies, parameter transferability~\cite{galda2023similarity, shaydulin2023parameter, falla2024graph} has been numerically investigated in the context of the Quantum Approximate Optimization Algorithm (QAOA)~\cite{farhi2000quantum}, a variational algorithm that exploits adiabatic evolution similar to QA.
Transferability allows the reuse of an optimized schedule across different problem instances, thus avoiding repeated schedule optimization.
If the parameters are transferable, the computational overhead of schedule optimization can be effectively eliminated.

We examined the transferability of the optimized schedule in the proposed method.
Figure~\ref{fig: transferability_bet_MWISs} shows the optimized schedules for different MWIS instances.
The variations among these schedules are negligible compared to the overall scale of the annealing schedule for both the transverse field and the Ising model.
This close similarity indicates a strong structural resemblance, implying that the optimized parameters are transferable across MWIS instances.
The dependence of the optimized schedule on the annealing time is provided in Appendix~\ref{sec: annealing_time_dependence_sche}.

The optimized annealing schedules decay during the first half of the anneal and oscillate while approaching positive values during the second half.
Near the initial time, adding $z$ field terms that do not commute with the transverse field enables the state to approach the computational basis rapidly.
As a result, the optimized schedule applies a relatively strong longitudinal field at early times.
This behavior is consistent with what has been described as a bang-bang regime in previous studies, and we interpret it as a closely related phenomenon~\cite{brady2021optimal,shirai2024post}.

As the evolution proceeds toward the final time, oscillatory behavior emerges in the optimized schedule. 
We attribute this to the need for the system to increasingly reflect the structure of the problem Hamiltonian near the end of the anneal.
In particular, the direction of the added longitudinal field differs from that of the target ground state.
In the present problem setting, the ground state consists of configurations with half of the spins pointing up and the other half pointing down.
The longitudinal fields required to align individual spins, therefore, compete with each other, leading to oscillations in the schedule.
Consequently, we expect the optimized longitudinal field to exhibit alternating positive and negative amplitudes as the evolution approaches the classical limit.

Figure~\ref{fig: transferability_bet_MWISs_TTS} compares the TTS between QA with a linear schedule and the proposed method.
The blue circles represent the proposed method with an individually optimized schedule $C(t)$, while the red crosses indicate the proposed method using a schedule $C(t)$ transferred from an arbitrary MWIS instance.
As shown in Fig.~\ref{fig: transferability_bet_MWISs_TTS}, the proposed method outperforms QA with a linear schedule even when using the transferred schedule.
These results demonstrate the feasibility of omitting schedule optimization costs for similar MWIS instances.

In Fig.~\ref{fig: transferability_bet_MWISs_TTS}, the transferred schedule yields shorter TTS than the optimized schedule in some instances.
This discrepancy arises from the slight variations in the optimized schedule shown in Fig.~\ref{fig: transferability_bet_MWISs}.
Because the optimization process in the proposed method converges differently across MWIS instances, the degree of convergence can vary even when the number of iterations and the learning rate are identical.

Understanding the underlying reason for the transferability is crucial for extending these results to other combinatorial optimization problems.
In MWIS instances, the energy gap becomes small near the end of QA.
According to the adiabatic theorem in quantum mechanics, the energy gap between the ground state and the first excited state during QA is a key property that strongly depends on the configuration of the problem Hamiltonian.

To analyze the impact of this energy gap structure, we construct a replaced energy-level Hamiltonian in which the Hamming distance between the ground state and the first excited state is set to $1$ while preserving the energy spectrum of the original Ising model for MWIS.
In contrast, the original Ising model for MWIS has a Hamming distance of $N$ between these states.
Using this replaced energy-level Hamiltonian, we examine the properties of schedule optimization and its transferability.
Figure~\ref{fig: replaced_Hamiltonian} shows schematic diagrams of the energy spectrum of the Ising Hamiltonian and the replaced energy-level Hamiltonian.

Figure~\ref{fig: transferability_replaced_Hamiltonian} shows the instantaneous ground-state population of the proposed method for hard MWIS instances under three different scheduling conditions: an optimized schedule, a transferred schedule, and a transferred schedule applied to the replaced energy-level Hamiltonian.
In Fig.~\ref{fig: transferability_replaced_Hamiltonian}, the ground-state population for the replaced energy-level Hamiltonian deviates significantly from that of the original Hamiltonian.

The proposed method with the optimized schedule and transferred schedule leverages adiabatic transitions during the time evolution, whereas the proposed method using the transferred schedule for the replaced energy-level Hamiltonian fails to effectively exploit diabatic transitions.
Consequently, schedule transferability does not hold between Ising models with the same energy spectrum but with different Hamming distances between the ground state and the first excited state.
This indicates that transferability is governed by the Hamming distance when diabatic transitions play a significant role in QA.

These findings suggest that the energy gap between the ground state and the first excited state during the annealing process alone cannot determine the performance of QA.

\section{\label{sec: discussion}Conclusion}
We investigated the effects of local diagonal catalysts on MWIS instances with a small energy gap near the final stages of QA.
In particular, we focused on the potential of local diagonal catalysts by optimizing the catalyst schedule for hard instances that are challenging to solve using conventional QA.
Our results demonstrate that the local diagonal catalyst significantly improves QA performance on MWIS.
The scaling analysis indicates that the proposed method achieves a quadratic speedup of the exponential scaling exponent.

Moreover, the dynamical properties of the system play a crucial role.
Diabatic transitions are essential for efficiently exploring the ground state in the proposed method.
We also examined the transferability of the catalyst schedule.
Our findings reveal that, in certain cases, the catalyst schedule optimized for complex MWIS instances can be transferred to other MWIS instances, particularly when QA with a linear schedule requires a long annealing time to reach the ground state.
This investigation is restricted to systems with the same number of spins; extending transferability to systems of different sizes remains a subject for future work.

By analyzing the replaced energy-level Hamiltonian, which preserves the energy spectrum of the original model while altering the Hamming distance, we revealed that the schedule optimization depends on both the energy levels and the Hamming distance from the ground state of the Ising model.
These results highlight the crucial role of the energy landscape of the Ising model in diabatic QA.
Therefore, the energy landscape transformation of the Ising model may provide an effective strategy to reduce the cost of scheduling optimization~\cite{fujii2022energy, fujii2023eigenvalue, kanai2024static}.

\begin{acknowledgments}
This work was partially supported by the Japan Society for the Promotion of Science (JSPS) KAKENHI (Grant Number JP23H05447), the Council for Science, Technology, and Innovation (CSTI) through the Cross-ministerial Strategic Innovation Promotion Program (SIP), ``Promoting the application of advanced quantum technology platforms to social issues'' (Funding agency: QST), Japan Science and Technology Agency (JST) (Grant Number JPMJPF2221). 
T.~H. was supported by JST SPRING, Grant Number JPMJSP2123, and NICT Quantum Camp 2023.
The computations in this work were partially performed using the facilities of the Supercomputer Center, the Institute for Solid State Physics, The University of Tokyo.
S.~T. wishes to express their gratitude to the World Premier International Research Center Initiative (WPI), MEXT, Japan, for their support of the Human Biology Microbiome-Quantum Research Center (Bio2Q).
\end{acknowledgments}
\section*{DATA AVAILABILITY}
The data that support the findings of this article are not publicly available upon publication because it is not technically feasible and/or the cost of preparing, depositing, and hosting the data would be prohibitive within the terms of this research project. The data are available from the authors upon reasonable request.
 
\appendix
\section{PROPOSED METHOD FOR SPIN GLASS\label{sec: SK}}
\begin{figure}
    \centering
    \includegraphics[scale=1,clip]{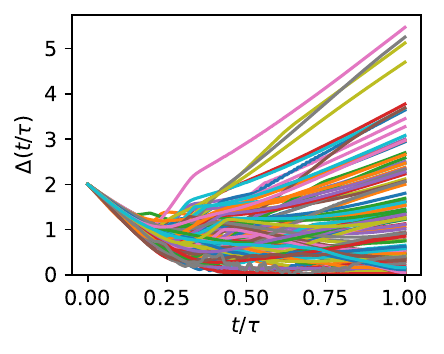}
    \caption{Energy gap $\Delta(t)$ between the ground state and the first excited state as a function of $t$ for different $100$ SK spin-glass problems with $N=8$.
    Different colors correspond to the energy gap for the different SK spin-glass problems.
    }
    \label{fig: SK_eg}
\end{figure}
\begin{figure}
    \centering
    \includegraphics[scale=1,clip]{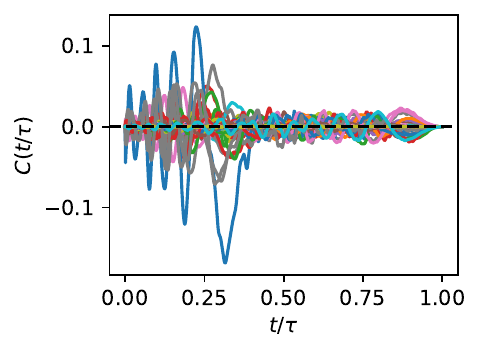}
    \caption{Optimized schedules for different $100$ SK spin-glass problems with $N=8$ at annealing time $\tau=100$.
    The black dashed line represents the zero reference, and different colors correspond to different SK spin-glass problems.
    }
    \label{fig: SK_sche}
\end{figure}
\begin{figure}
    \centering
    \includegraphics[scale=1,clip]{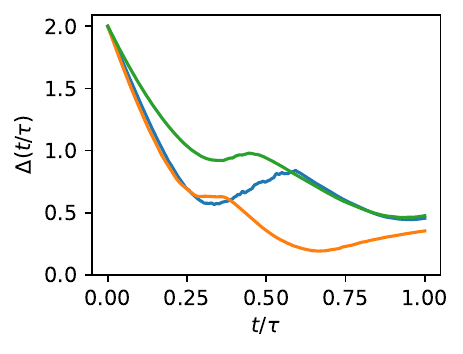}
    \caption{Energy gap $\Delta(t)$ between the ground state and the first excited state as a function of $t$ for three SK spin-glass problems with $N=8$.
    Different colors correspond to the energy gap for the different SK spin-glass problems.
    }
    \label{fig: SK_sorted_eg}
\end{figure}
\begin{figure}
    \centering
    \includegraphics[scale=1,clip]{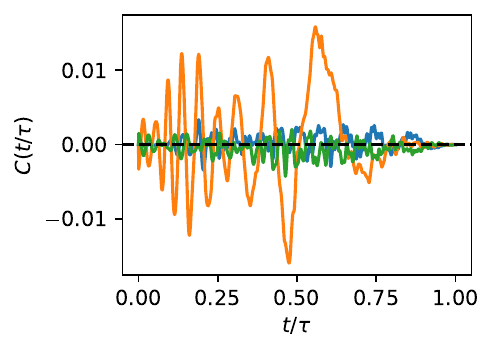}
    \caption{Optimized schedules for three SK spin-glass problems with $N=8$ at annealing time $\tau=100$.
    The black dashed line represents the zero reference, and different colors correspond to different SK spin-glass problems.
    }
    \label{fig: SK_sorted_sche}
\end{figure}
\begin{figure}
    \centering
    \includegraphics[scale=1,clip]{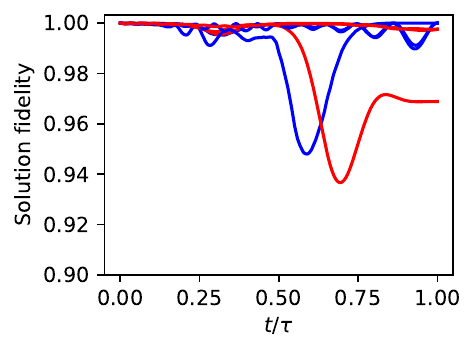}
    \caption{Ground-state population during the time evolution for three SK spin-glass problems with $N=8$.
    The annealing time is set to $\tau=100$.
    The blue and red solid lines show the ground-state populations for the proposed method and the linear-schedule QA, respectively.
    The different lines correspond to the different SK spin-glass problems.
    }
    \label{fig: ins_prob}
\end{figure}
In this section, we examined the performance of the proposed method on the Sherrington–Kirkpatrick (SK) spin-glass problem.
The size of the SK spin-glass problems is $N=8$ with both longitudinal magnetic fields and interactions randomly generated from a normal distribution with zero mean and unit variance.
We generated $100$ such SK instances and investigated the properties of the proposed method.
Here, we set the annealing time to $\tau=100$, which is motivated by the observation that this annealing time yields dynamics close to adiabatic evolution.

Figure~\ref{fig: SK_eg} shows the energy gaps obtained when conventional QA is applied to the $100$ generated SK instances.
As seen from Fig.~\ref{fig: SK_eg}, the energy-gap structure varies significantly from instance to instance in this problem setting.
Figure~\ref{fig: SK_sche} shows the annealing schedules of the catalyst term optimized by the proposed method for these instances.
The optimized schedules differ substantially across problem instances, indicating that schedule transferability is not observed for general SK models with varying problem structures and energy-gap magnitudes.
At first glance, this result may appear unfavorable, as it suggests limited effectiveness of the proposed approach.
However, problem instances with large energy gaps are intrinsically easy to solve using conventional quantum annealing.
Since the $100$ generated instances exhibit a wide range of energy-gap sizes and corresponding degrees of difficulty, this observation should not be interpreted pessimistically.

We therefore focus on the three instances among the $100$ that exhibit the smallest minimum energy gaps.
Figure~\ref{fig: SK_sorted_eg} shows the energy gaps for these three instances under conventional quantum annealing.
Among them, two instances display similar gap behaviors and have comparable minimum gap values.
Figure~\ref{fig: SK_sorted_sche} shows the optimized catalyst annealing schedules obtained for these three instances using the proposed method.
As seen in Fig.~\ref{fig: SK_sorted_sche}, the two instances with similar energy-gap structures also exhibit similar behaviors in the optimized catalyst schedules.
Figure~\ref{fig: ins_prob} shows the corresponding energy-level populations.
By introducing the catalyst term, nonadiabatic transitions are exploited, leading to a significant enhancement of the ground-state population at the final time.

These results indicate that the proposed method is particularly effective for problem instances in which purely adiabatic time evolution is difficult.
The underlying mechanism for the performance improvement can be understood as follows.
The system initially follows an approximately adiabatic evolution; however, near regions where the energy gap between the ground state and the first excited state becomes small during the annealing process, nonadiabatic transitions are intentionally induced.
The mechanism allows population transfer from the first excited state back to the ground state when the gap is small, enabling performance enhancement through dynamics that actively exploit nonadiabatic transitions.
These results are consistent with the results of the proposed method on MWIS instances.

\section{\label{sec: Derivation_optsche}DERIVATION OF THE GRADIENT OF THE SCHEDULE}
In this appendix, we derive the gradient of the schedule by using optimal control theory.
First, we define the control problem as follows:
\begin{align}
    \label{eq: control_problem_appendix}
    \mathcal{J} \equiv\bra{\psi(\tau)}\mathcal{H}_\mathrm{p}\ket{\psi(\tau)}.
\end{align}
Equation \eqref{eq: control_problem_appendix} indicates the expected energy at the final time of QA.
Therefore, the schedule optimization aims to reduce energy.
Here, we introduce the time evolution operator as follows:
\begin{align}
    \label{eq: time_dependent_time_evolution_operator}
    U(t,0)=\mathcal{T}\left[\exp \left( -\mathrm{i} \int^t_0 \mathcal{H}(t^{\prime})dt^{\prime}\right)\right],
\end{align}
where $\mathcal{T}$ represents time-ordered product.
Therefore, the control problem defined in Eq.~\eqref{eq: control_problem_appendix} is expressed using time evolution operator written in Eq.~\eqref{eq: time_dependent_time_evolution_operator} as follows:
\begin{align}
    \label{eq: objective_control_x_0}
    \mathcal{J}&=\bra{\psi(0)}U^\dag(\tau,0)\mathcal{H}_\mathrm{p}U(\tau,0)\ket{\psi(0)}.
\end{align}
Here, when we assume the time difference is negligible, the Hamiltonian can be considered as $H(t)$ from $t$ to $t+\Delta t$.
The time evolution operator is approximated as follows:
\begin{align}
    \label{eq: time_independent_time_evolution_operator}
    U(t+\Delta t,t)\approx \exp\left(-\mathrm{i}\Delta t\mathcal{H}(t)\right).
\end{align}
By using Eq.~\eqref{eq: time_independent_time_evolution_operator}, the integration in Eq.~\eqref{eq: time_dependent_time_evolution_operator} can be written as
\begin{widetext}
\begin{align}
    \label{eq: time_evolution_state}
    U(\tau,0)=&\lim_{n\rightarrow \infty}U\left(\tau,\tau-\frac{\tau}{n}\right)U\left(\tau -\frac{\tau}{n},\tau-2\frac{\tau}{n}\right)\cdots U\left(2\frac{\tau}{n},\frac{\tau}{n}\right)U\left(\frac{\tau}{n},0\right)\nonumber\\
    =&\lim_{\Delta t\rightarrow 0}U(\tau,\tau-\Delta t )U(\tau -\Delta t,\tau-2\Delta t)\cdots U(2\Delta t,\Delta t)U(\Delta t,0),
\end{align}
where we set $\tau/n =\Delta t$.
Here, we use the short-time approximation for a small time step $\Delta t$.
The differentiation of the Eq.~\eqref{eq: time_evolution_state} is 
\begin{align}
    \frac{\partial U(\tau,0)}{\partial C(t)}
    &=\frac{\partial}{\partial C(t)} \left[U(\tau,\tau -\Delta t)U(\tau -\Delta t,\tau - 2\Delta t)\dots U(2\Delta t,\Delta t)U(\Delta t,0))\right]\nonumber\\
    &=\frac{\partial}{\partial C(t)} \left[U(\tau,\tau -\Delta t)\right]U(\tau -\Delta t,\tau - 2\Delta t)\dots U(2\Delta t,\Delta t)U(\Delta t,0)\nonumber\\
    &+ U(\tau,\tau -\Delta t)\frac{\partial}{\partial C(t)}\left[U(\tau -\Delta t,\tau -2\Delta t)\right]\dots U(2\Delta t,\Delta t)U(\Delta t,0)\nonumber\\
    &+\cdots\\
    &+ U(\tau,\tau-\Delta t)U(\tau -\Delta t,\tau -2\Delta t)\dots \frac{\partial}{\partial C(t)}\left[U(2\Delta t,\Delta t)\right]U(\Delta t,0)\nonumber\\
    &+ U(\tau,\tau-\Delta t)U(\tau -\Delta t,\tau -2\Delta t)\dots U(2\Delta t,\Delta t)\frac{\partial}{\partial C(t)}\left[U(\Delta t,0)\right]\nonumber\\
    &=\sum_{m=1}^{n}\left\{\left[\Pi_{i=m+1}^{n} U(i\Delta t,(i-1)\Delta t) \right]\frac{\partial}{\partial C(t)}U(m\Delta t,(m-1)\Delta t)\left[\Pi_{i=1}^{m-1}U((i+1)\Delta t,i\Delta t)\right]\right\}.\nonumber\\
    \label{eq: time_evolution_state_differential}
\end{align}
Here, the empty product is unity.
We use the Duhamel formula, the Eq.~\eqref{eq: time_evolution_state_differential} is expressed as follows: 
\begin{align}
    \label{eq: U_grad}
    \frac{\partial U(m\Delta t,(m-1)\Delta t)}{\partial C(t)}&=-\mathrm{i}\Delta t\int^1_0 \exp\left(-(1-k)\mathrm{i}\Delta t \mathcal{H}(t)\right)\frac{\partial\mathcal{H}(t)}{\partial C(t)}\exp\left(-k\mathrm{i}\Delta t\mathcal{H}(t)\right)dk.
\end{align}
\end{widetext}
Here, by taking the limit $\Delta t\rightarrow 0 ~(n\rightarrow\infty)$, and retaining only the first-order accuracy,  
\begin{align}
    \exp(-\mathrm{i}(1-k)\Delta t\mathcal{H}(t))=1+\mathcal{O}(\Delta t),\\
    \exp(-\mathrm{i}k\Delta t\mathcal{H}(t))=1+\mathcal{O}(\Delta t),
\end{align}
is obtained.
This, in turn, leads to
\begin{align}    
    \frac{\partial U(m\Delta t,(m-1)\Delta t)}{\partial C(t)}&\overset{n\rightarrow\infty}{=} -\mathrm{i}\Delta t\frac{\partial \mathcal{H}(t)}{\partial C(t)}.
\end{align}
Therefore, the differentiation of the time evolution operator from $0$ to $\tau$ is as follows:
\begin{align}
    \label{eq: U_grad2}
    \frac{\partial U(\tau,0)}{\partial C(t)}
    &=-\mathrm{i}U(\tau,t)\frac{\partial \mathcal{H}(t)}{\partial C(t)}U(t,0),
\end{align}

The differentiation of the gradient of the control problem shown in Eq.~\eqref{eq: control_problem_appendix} is calculated as follows:  
\begin{align}
    \label{eq: gradient_J}
    \frac{\partial \mathcal{J}}{\partial C(t)}&=\frac{\partial}{\partial C(t)}\left(\bra{\psi(\tau)}\mathcal{H}_\mathrm{p}\ket{\psi(\tau)}\right)\nonumber\\
    &=\frac{\partial \bra{\psi(\tau)}}{\partial C(t)}\mathcal{H}_\mathrm{p}\ket{\psi(\tau)}+\bra{\psi(\tau)}\mathcal{H}_\mathrm{p}\frac{\partial \ket{\psi(\tau)}}{\partial C(t)},
\end{align}
Here, the conjugate of the first term in Eq.~\eqref{eq: gradient_J} satisfies the equation: 
\begin{align}
    \label{eq: cc_gradient_J}
    \left(\frac{\partial \bra{\psi(\tau)}}{\partial C(t)}\mathcal{H}_\mathrm{p}\ket{\psi(\tau)}\right)^\dag=\bra{\psi(\tau)}\mathcal{H}_\mathrm{p}\frac{\partial \ket{\psi(\tau)}}{\partial C(t)}.
\end{align}
Therefore, note that the first term and the second term in Eq.~\eqref{eq: gradient_J} are conjugate to each other, and Eq.~\eqref{eq: gradient_J} is as follows:
\begin{align}
    \label{eq: gradient_J_2}
    \frac{\partial \mathcal{J}}{\partial C(t)}&=2\mathrm{Re}\left(\bra{\psi(\tau)}\mathcal{H}_\mathrm{p}\frac{\partial \ket{\psi(\tau)}}{\partial C(t)}\right)\nonumber\\
    &=2\mathrm{Re}\left(\bra{\psi(\tau)}\mathcal{H}_\mathrm{p}\frac{\partial U(\tau,0)}{\partial C(t)}\ket{\psi(0)}\right),
\end{align}
where $\mathrm{Re}(\cdot)$ indicate the real part of the complex number.
As a result, the gradient of the annealing schedule is expressed as follows:
\begin{align}
    \label{eq: Gradient_J_u}
    \frac{\partial \mathcal{J}}{\partial C(t)}&=2\mathrm{Re}\left(\bra{\psi(\tau)}\mathcal{H}_\mathrm{p}U(\tau,t)\frac{\partial \mathcal{H}(t)}{\partial C(t)}U(t,0)\ket{\psi(0)}\right)\nonumber\\
    &=2\mathrm{Re}\left(\bra{k(t)}\mathcal{H}_\mathrm{p}\frac{\partial \mathcal{H}(t)}{\partial C(t)}\ket{\psi(t)}\right),
\end{align}
where $\ket{k(t)}$ satisfy the equation:
\begin{align}
    \label{eq: k_vector}
    \ket{k(t)}=\mathcal{H}_\mathrm{p}\ket{\psi(t)},
\end{align}
Assigning $\mathcal{H}$ in Eq.~\eqref{eq: H_catalyst} to Eq.~\eqref{eq: Gradient_J_u} results in the equation:
\begin{align}
    \label{eq: QA_gradient}
    \frac{\partial \mathcal{J}}{\partial C(t)}&=2\mathrm{Im}\left(\bra{k(t)}\mathcal{H}_\mathrm{catalyst}\ket{\psi(t)}\right),
\end{align}
where $\mathrm{Im}(\cdot)$ indicates the imaginary part of the complex number.
In this study, the gradient shown in Eq.~\eqref{eq: QA_gradient} is calculated by solving the Schr\"{o}dinger equation.
\begin{figure*}[t]
    \begin{tabular}{cc}
    \begin{minipage}[c]{0.5\hsize}
    \includegraphics[clip,scale=1.0]{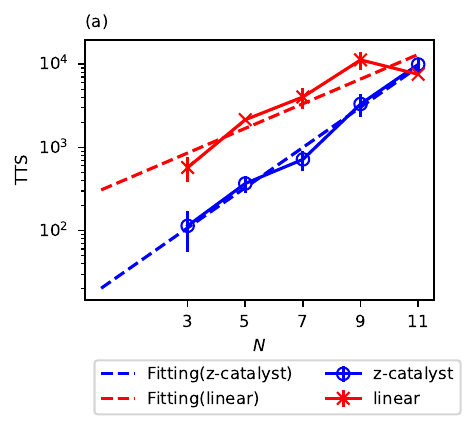}
    \end{minipage}
    \begin{minipage}[c]{0.5\hsize}
    \includegraphics[clip,scale=1.0]{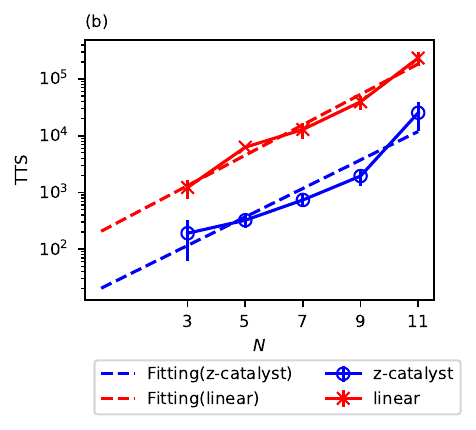}
    \end{minipage}\\
    \begin{minipage}[c]{0.5\hsize}
    \includegraphics[clip,scale=1.0]{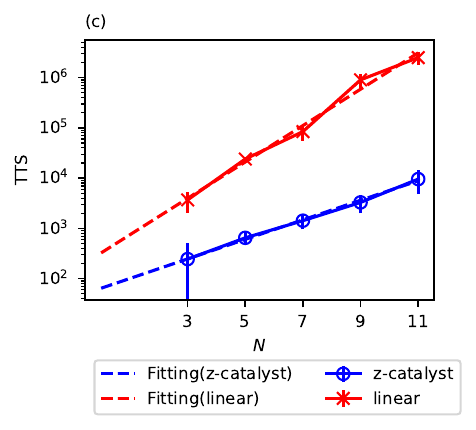}
    \end{minipage}
    \begin{minipage}[c]{0.5\hsize}
    \includegraphics[clip,scale=1.0]{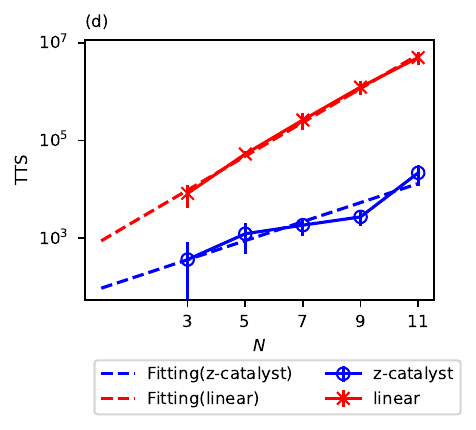}
    \end{minipage}
    \end{tabular}
    \caption{Scaling of TTS with system size for hard MWIS instances.
    The vertical axis is logarithmic.
    Points show mean TTS values, error bars denote standard deviations, and dashed lines indicate least-squares fits.
    Panels correspond to different annealing times: (a) $\tau=16$, (b) $\tau=32$, (c) $\tau=100$, and (d) $\tau=256$.}
    \label{fig: TTSscaling_tau_dependence}
\end{figure*}

\section{\label{sec: annealing_time_dependence}ANNEALING TIME DEPENDENCE OF TIME TO SOLUTION}
\begin{table*}[t]
    \caption{Fitting functions for TTS obtained using the least-squares method. 
    $C_1$ and $C_2$ represent constant parameters.}
    \centering
    \begin{tabular*}{\textwidth}{@{\extracolsep{\fill}}lcc}
        \hline
        \hline
        Annealing time &Linear schedule & Proposed method \\ \hline
        $\tau=16$& $T = C_1\exp(0.34N\pm0.0826)$ & $T = C_2\exp(0.55N\pm0.0377)$\\ \hline
        $\tau=32$& $T = C_1\exp(0.62N\pm0.0519)$ & $T = C_2\exp(0.58N\pm0.1124)$\\ \hline
        $\tau=100$& $T = C_1\exp(0.83N\pm0.0559)$ & $T = C_2\exp(0.45N\pm0.0132)$\\ \hline
        $\tau=256$& $T = C_1\exp(0.80N\pm0.0221)$ & $T = C_2\exp(0.45N\pm0.0821)$\\ \hline\hline
    \end{tabular*}
    \label{tab: TTS_sizescaling_appendix}
\end{table*}
\begin{figure*}[t]
    \begin{tabular}{cc}
    \begin{minipage}[c]{0.5\hsize}
    \includegraphics[clip,scale=1.0]{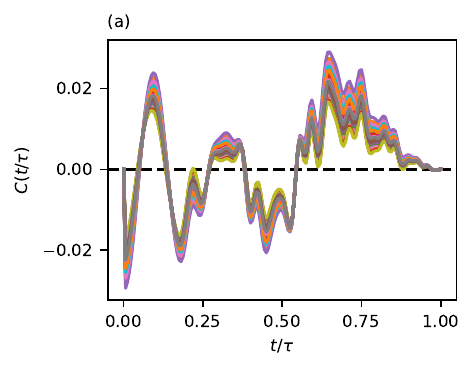}
    \end{minipage}
    \begin{minipage}[c]{0.5\hsize}
    \includegraphics[clip,scale=1.0]{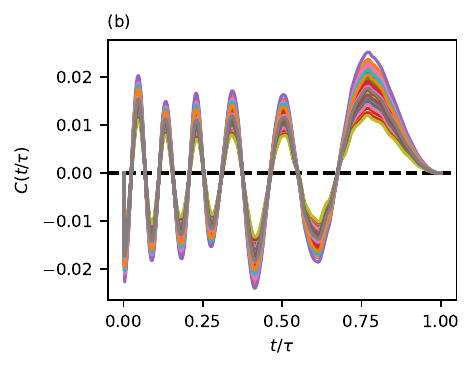}
    \end{minipage}\\
    \begin{minipage}[c]{0.5\hsize}
    \includegraphics[clip,scale=1.0]{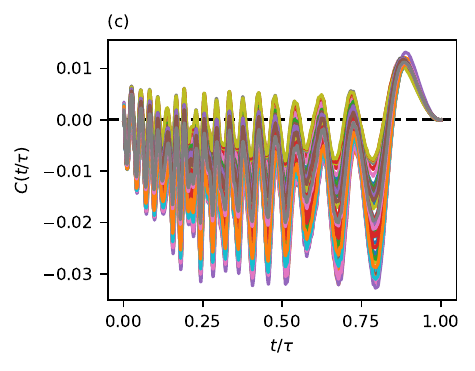}
    \end{minipage}
    \begin{minipage}[c]{0.5\hsize}
    \includegraphics[clip,scale=1.0]{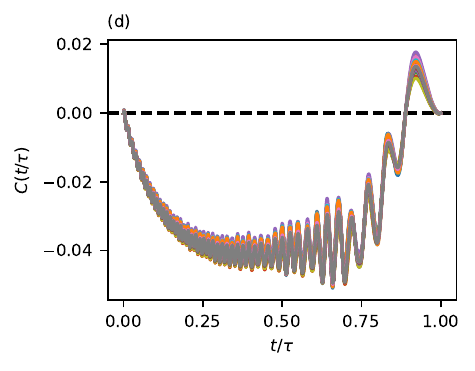}
    \end{minipage}
    \end{tabular}
    \caption{Optimized schedule for hard MWIS instances with $N = 11$.
    The black dashed line represents the reference line at zero. 
    Different colors indicate the results for different MWIS instances.
    Panels show results for different annealing durations: (a) $\tau=16$, (b) $\tau=32$, (c) $\tau=100$, and (d) $\tau=256$.}
    \label{fig: schedule_tau_dependence}
\end{figure*}
Figure~\ref{fig: TTSscaling_tau_dependence} shows the annealing-time dependence of TTS, and Table~\ref{tab: TTS_sizescaling_appendix} summarizes the fitting functions obtained using the least-squares method.
Figure~\ref{fig: TTSscaling_tau_dependence} (a), for $\tau=16$ and $N=11$, the linear-schedule QA achieves shorter TTS than the proposed method.
In this condition, the system remains in its initial state, with the QA of a linear schedule, similar to uniform random sampling.
Also, the scaling of the TTS with $N$ deviates from the exponential fitting function.
According to the comparison of Fig.~\ref{fig: TTSscaling_tau_dependence} (a)-(d) and Table~\ref{tab: TTS_sizescaling_appendix}, performance gains from the proposed method become more pronounced as $\tau$ increases.
These results indicate the proposed method is effective only for sufficiently long $\tau$.
When $\tau$ is too short to reach the ground state, the system moves to the local minimum to reduce the energy instead.
Since the proposed method minimizes the final-time energy [Eq.~\eqref{eq: control_problem}], it still yields lower energies than conventional QA.
However, this mechanism explains why the proposed method exhibits worse TTS in the short-$\tau$ regime.
In the case of a linear schedule, Fig.~\ref{fig: TTSscaling_tau_dependence} shows that the TTS increases as the annealing time becomes longer.
This phenomenon arises when the annealing time is extended but remains insufficient to overcome the small energy gap.
In general, increasing the annealing time is expected to enhance adiabaticity, thereby raising the ground-state population.
However, in the present problem setting, the energy gap is so small that nonadiabatic transitions cannot be avoided even with longer annealing times.
Consequently, the ground-state population does not increase with annealing time, leading to a longer TTS.
This behavior can also be observed in the proposed method, reflecting the trade-off between extending the annealing time per run and performing multiple repetitions of QA. 
It indicates that the improvement in TTS obtained by further increasing the annealing time cannot surpass the improvement achieved by executing multiple QA runs.

\section{\label{sec: annealing_time_dependence_sche}ANNEALING TIME DEPENDENCE OF THE OPTIMIZED ANNEALING SCHEDULE}
Figure~\ref{fig: schedule_tau_dependence} shows the dependence of the optimized annealing schedule on the annealing time in the proposed method.
Figures~\ref{fig: schedule_tau_dependence} (a)-(d) show that the optimized annealing schedule for different MWIS with $N=11$ when $\tau=16.0$, $\tau=32.0$, $\tau=100.0$ and $\tau=256.0$, respectively.
When the annealing time is sufficiently long, differences in convergence are minor.
Across different annealing times, the variations among these schedules remain negligible compared with the overall scale of the annealing schedules for both the transverse field and the Ising model.
Therefore, the optimized schedule transferabilities are confirmed.

\bibliographystyle{apsrev4-2}
\bibliography{03-journal-TomohiroHattori}
\end{document}